# Improving Deep Speech Denoising by Noisy2Noisy Signal Mapping

N. Alamdari, *Student Member, IEEE*, A. Azarang, *Member, IEEE*, N. Kehtarnavaz, *Fellow, IEEE*

*Abstract*—**Existing deep learning-based speech denoising approaches require clean speech signals to be available for training. This paper presents a deep learning-based approach to improve speech denoising in real-world audio environments by not requiring the availability of clean speech signals in a self-supervised manner. A fully convolutional neural network is trained by using two noisy realizations of the same speech signal, one used as the input and the other as the output of the network. Extensive experimentations are conducted to show the superiority of the developed deep speech denoising approach over the conventional supervised deep speech denoising approach based on four commonly used performance metrics and also based on actual field-testing outcomes.**

## I. Introduction

SPEECH denoising is extensively studied in the literature. The existing speech denoising methods can be categorized into two main groups: (1) *Conventional methods* – These methods involve estimating the noise to achieve denoising. Examples of these methods are spectral subtraction [1] and Wiener filtering [2, 3]. (2) *Deep learning-based methods* – These more recent methods attempt to model the nonlinear relationship between noisy and clean speech signals via a deep neural network (DNN), e.g. [4-8]. These methods have allowed dealing with non-stationary audio environments [9] and can be further divided into two categories: direct mapping (mapping-based) methods, e.g. [10-12], which mostly use the log-power spectra as the input and output of a DNN and masking-based methods, e.g. [13-15], which estimate a mask to carry out denoising.

A major assumption made in existing deep learning-based methods is the availability of clean speech signals to conduct training. In practice, the problem of speech denoising is more challenging due to the fact that clean speech signals are not known or available when operating in the field or in real-world audio environments. In addition, the generalization capability of current DNN-based approaches is limited for unseen speakers and varying signal-to-noise ratios. In [16], an attempt was made to perform denoising without using clean speech signals. The assumption made in [16] was to have simultaneous access to both noise-only and speech+noise signals. In practice, such simultaneous access to noise only and speech+noise signals is very difficult to achieve and only one of these signals is available at any given time.

In this work, a deep speech denoising approach is introduced to ease the major assumption of availability of clean speech signals in the existing deep speech denoising solutions. As a result, the introduced approach can be deployed in real-world audio environments or in the field in which clean speech signals are not available. Not requiring to have clean speech signals is what differentiates this paper from the existing deep speech denoising papers.

The rest of the paper is organized as follows: In section II, the introduced deep speech denoising approach and its implementation aspects are presented. In Section III, an overview of the datasets examined is provided. A comprehensive set of experimentations and their results are then reported in Section IV. Finally, the paper is concluded in Section V.

## II. Developed Deep Speech Denoising Approach

The denoising approach developed in this paper builds upon the commonly practiced denoising approach of supervised training. A deep neural network is first initialized based on the public domain datasets for which clean speech signals are available. Then, the network is further trained in a self-supervised manner based on only noisy speech data or in the absence of clean speech signals.

As input to a deep neural network, various audio representation schemes have been utilized. Among them, mel frequency cepstral coefficient (MFCC) [17] has been widely used. Also, log-mel spectrogram coefficient (MFSC) has been used by omitting the discrete cosine transform (DCT) compression from the MFCC computation. These frequency-domain representations focus on the magnitude spectrum and the phase spectrum is often left unprocessed [18]. Similar to recent studies [19-22], raw waveform or time-domain signals are considered here as input to a deep neural network instead of frequency-domain representations.

In supervised deep learning-based speech denoising, a network is trained to perform noise reduction by considering clean speech signals as its output or target signals. A speech+noise mixture or noisy speech signal $y(t)$ can be expressed as

$$y(t) = x(t) + n(t) \qquad (1)$$

where $x(t)$ and $n(t)$ denote clean speech and additive noise signals, respectively. A reasonable assumption which is often made is that clean speech and noise signals are uncorrelated and noise is zero mean [23]. A network is then trained and used to generate an estimate of the denoised speech signal $\hat{x}_i = f(y_i; \Theta)$ based on the noisy speech signal $y_i$ as its input. The index $i$ is used here to indicate signal frames.

After the initialization of the network in a supervised manner, the self-supervised training is conducted by considering noisy speech signals as both the input and the output of the network *without knowing clean speech signals*. In contrast to supervised denoising, this self-supervised training makes it more suitable for field deployment as in the field clean speech signals are not available.

As illustrated in Fig. 1, in the supervised speech denoising approach (labeled SSD here), training pairs $(y_i, x_i)$ are used to minimize the network loss function in which $y_i$ is the noisy speech input frame and $x_i$ is the corresponding clean speech target frame. The weights of the network are obtained by solving the following optimization problem

$$arg \min_{\Theta} \sum_i \mathcal{L}(f(y_i; \Theta) = \hat{x}_i, x_i) \qquad (2)$$

where $\mathcal{L}(.)$ denotes a loss function, normally the mean squared error (MSE) function, and $\Theta$ denotes the network parameters or connection weights.

In the self-supervised approach, see Fig. 1, two noisy realizations of clean speech signals are used during training. In other words, the input and the target are considered to be two noisy versions of the same speech signal instead of the target being the clean speech signal as in the supervised approach. This means that the network is trained to solve the following optimization problem instead of the optimization problem in Eq. (2)

$$arg \min_{\Theta} \sum_i \mathcal{L}(f(y_i; \Theta) = \hat{x}_i, y'_i) \qquad (3)$$

In Eq. (3), $y'_i$ denotes another noisy realization of the same speech signal instead of the clean speech signal $x_i$. In other words, training pairs $(y, y')$ consisting of two noisy realizations of the same speech signal are used to train the network, $y_i = x_i + n_i$ and $y'_i = x_i + n'_i$.

This approach is inspired from the image denoising work reported in [24], named Nosie2Noise, where image denoising was achieved without using any clean image data. During training, if clean speech target signals are replaced with noisy speech signals with the expected value being equal to clean speech signals ($\mathbb{E}[y_i] \cong x_i$), the final weights of the network would remain more or less the same provided that two assumptions are met as pointed out in [24]. The first assumption is that the noise to be zero mean which is a reasonable assumption to make noting that in practice noise is often observed to have zero mean. The second assumption is that the two noisy signals $y$ and $y'$ need to be decorrelated or ideally uncorrelated. This assumption is met here by using a mid-side microphone during field training which is discussed next.

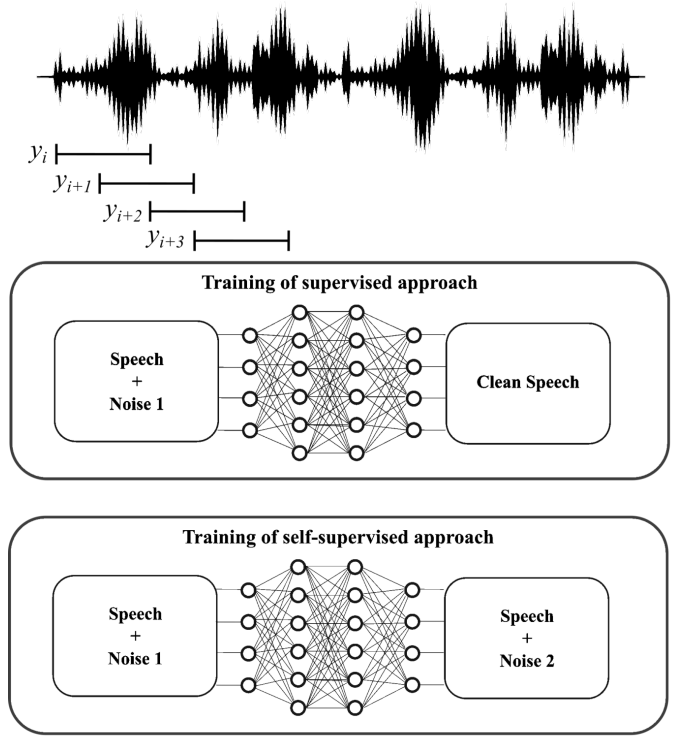

Fig. 1. Illustration of the training difference between the supervised deep speech denoising (top) and the self-supervised deep speech denoising (bottom).

*A. Mid-Side Microphone Signals*

A mid-side microphone is a stereo microphone whose two signals are generated based on the difference in loudness instead of time-delay. The relationship between the right-left channels and the mid-side microphone signals is given by [25]:

$$\begin{aligned} y_L &= y_M + y_S \\ y_R &= y_M - y_S \end{aligned} \qquad (4)$$

where the mid microphone $y_M$ is usually faced toward the speech source signal and the bi-directional side microphone $y_S$ with 90-degree rotation with respect to mid, captures the background noise signal. Noise signals normally arrive at the side-microphone along different paths and they have similar magnitudes but different phases, that is in the frequency-domain $|N_R| = |N_L|$ and $N_R = e^{j\phi} N_L$ [25]. As shown in Eq. (4), for the left channel, the side signal is added to the mid signal with positive polarity; and for the right channel, the side signal is added to the mid signal with negative polarity. As a result, the right and left signals become decorrelated [26].

Note that a low-quality separation of speech and background noise signals can be achieved by summing the right and left channels. Fig. 2 shows sample speech signals that are captured by a mid-side stereo microphone (Zoom iO7) in an actual audio environment of cafeteria babble noise. The first two signals from the top show the right $y_R$ and the left $y_L$ channels, respectively. The other two signals correspond to the side signal $y_S$ and the mid signal $y_M$, which are obtained from Eq. (4).

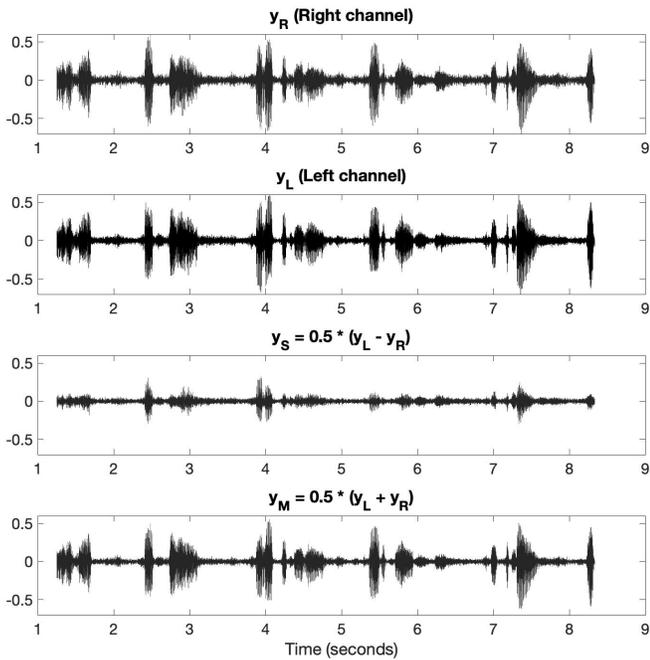

Fig. 2. Sample field audio signals captured by a mid-side stereo microphone.

### B. Architecture of Deep Neural Network

The training of the speech denoiser network is performed in a frame-based manner. As illustrated in Fig. 1, noisy speech signals are partitioned into 20ms frames through a Hanning window with 50% overlap between adjacent frames. A deep neural network is then used to obtain the denoised speech signal by using $y_i = x_i + n_i$ as its input and $y'_i = x_i + n'_i$ as its target, where $y_i$ and $y'_i$ denote two noisy realizations of the same speech captured by a mid-side stereo microphone.

Next, the architecture of the deep neural network used is mentioned. The network considered is a fully convolutional neural network (FCNN). The FCNN architecture is similar to a conventional convolutional neural network (CNN) architecture. The only difference is that the fully-connected layers are omitted in FCNN. As noted in [22], FCNNs can model the temporal attributes of time series data using 1D convolution layers. The FCNN architecture used here is similar to the one described in [22]. This architecture incorporates 6 convolution layers. The number of filters and filter size are 55 and (30,1) for the first through the fifth convolution layers, respectively. For the last convolution layer, only one filter of size (1,1) is used followed by the hyperbolic tangent activation function.

The training is speeded up by using the batch normalization in [27] and the Leaky Rectified Linear Units (LeakyReLU) activation function is applied after each convolution layer except for the last layer. In this architecture, MSE is used as the loss function. To train the network, the Adam optimization algorithm described in [28] is used, and the size of the minibatch and the initial learning rate are set to 128 and 0.0004, respectively. The training is normally performed for 25 epochs. As mentioned earlier, the main difference between FCNN and CNN is that there is no fully-connected layer in the output of FCNN. Also, the max-pooling layers are removed. As a result, in FCNN, the output frame depends only on the neighboring input frames. A depiction of the FCNN architecture is provided in Fig. 3.

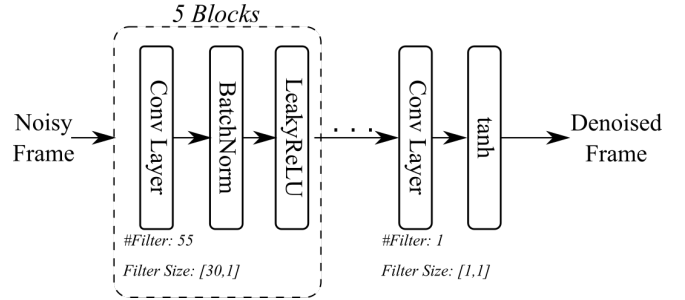

Fig. 3. FCNN architecture.

### C. Single Channel Operation or Testing

It needs to be noted that only for training of the self-supervised network, two channels are used. During the actual operation or testing of the developed deep speech denoising approach, only a single channel is used to feed noisy speech signal frames into the trained deep neural network with the output being denoised speech frames.

Fig. 4 shows the block diagram of the implementation pipeline for field-deployment of the developed deep speech denoising solution. Captured audio signals go through audio framing and averaging to provide the input to a voice activity detection (VAD) module as described in [29]. This module separates noisy speech frames from the absence of speech or from noise-only frames. Noise-only frames are then used to identify different noise types via an unsupervised noise classifier such as the one described in [30]. The identification of different noise types allows a bank of fully convolutional neural networks (FCNNs) to be trained, each network for a particular noise type with the number of noise types capped by the user. The selection of a FCNN model for a particular noise type among the bank of the FCNN models is done by the unsupervised noise classifier.

When a new noise type $n_{new}$ is encountered by the unsupervised noise classifier, a new FCNN can get trained based on speech signals that are corrupted by $n_{new}$. Then, this new network gets added to the network bank. In other words, among all the candidate networks, only one FCNN network will be selected based on the noise type during actual operation or testing in the field. The right part of Fig. 4 shows the process of training and operation/testing when a network is selected based on an identified noise type. Once the training switch is activated, see Fig. 4, noisy speech frames that belong to the right and left channels are used as the input and the target of the selected network for the self-supervised training. When the training switch is turned to the *off* mode, the network goes to the testing or operation mode. The denoising outcome is then played back through the speaker. It is important to note that the introduced approach uses the conventional supervised training as its initial condition. That is why it is labeled here as hybrid speech denoising (HSD).

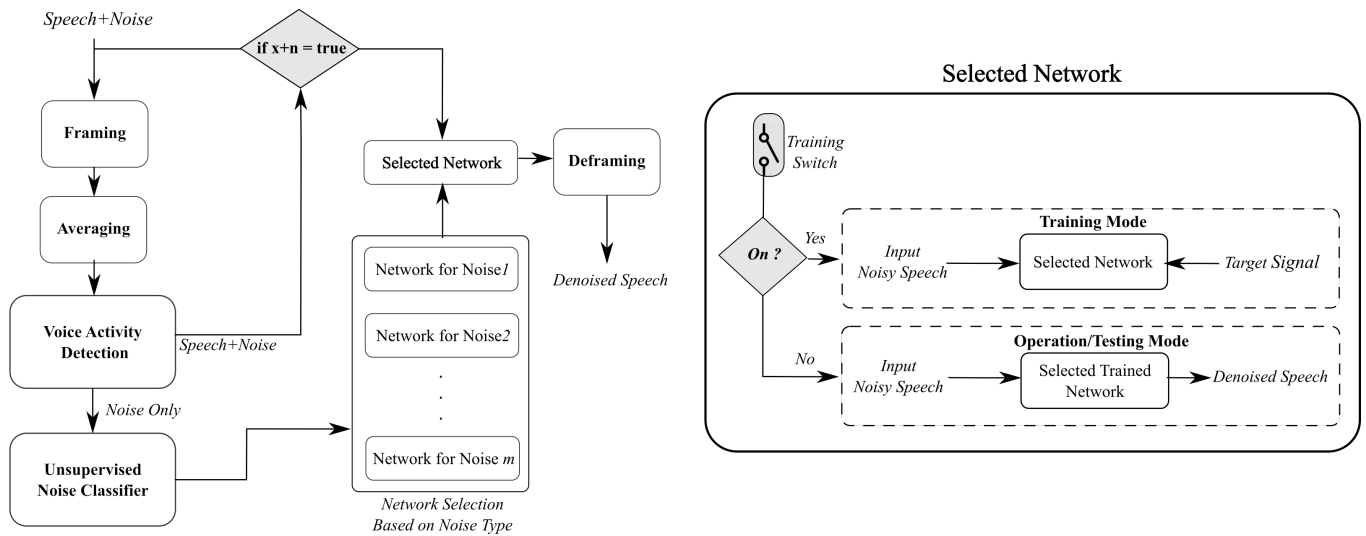

Fig. 4. Implementation pipeline of the developed deep speech denoising.

## III. PUBLIC DOMAIN DATASETS

Three widely used public domain speech datasets, IEEE [31], TIMIT [32], and VCTK [33] are considered for the experimentations reported in the next section. The IEEE Corpus consists of 3600 speech audio files by 20 speakers (10 females and 10 males) in which each file is about 2 seconds long. The speakers are from two American English regions of the Pacific Northwest (PN) and the Northern Cities (NC) reading the IEEE "Harvard" sentences. The TIMIT Corpus (Acoustic-Phonetic Continuous Speech) consists of 630 speakers from eight major American English dialects, each reading ten sentences. The VCTK Corpus (Centre for Speech Technology Voice Cloning Toolkit) contains audio files of 109 English native speakers with different accents. About 400 sentences are read by each speaker.

The above clean speech signals are corrupted by the noise signals from the UrbanSound8K dataset [34] consisting of various noise files each lasting 4 seconds. In this work, based on the urban sound taxonomy described in [34], the following four most commonly encountered noise signals are considered: (i) babble (e.g., restaurant, cafeteria), (ii) wind, (iii) engine, and (iv) driving car. All noisy speech files are sampled at 48 kHz and normalized to have absolute unit maximum.

## IV. EXPERIMENTAL RESULTS AND DISCUSSION

Four commonly used objective performance metrics were considered to assess the effectiveness of the developed HSD approach. These metrics include: perceptual evaluation of speech quality (PESQ) [35], short-time objective intelligibility (STOI) [36], segmental SNR (SSNR) [37], and log spectral distance (LSD). For the PESQ and STOI metrics, the ranges are [-0.5, 4.5] and [0, 1], respectively, in which the upper bound of the ranges corresponds to the ideal values. Higher SSNR means better performance and in case of LSD, the ideal value is 0.

Three sets of experiments were conducted. The first two sets of experiments were performed using the public domain datasets to compare the HSD approach with the conventional SSD approach. To perform the comparison in a fair manner, the same procedure depicted in Fig. 4 was used for the SSD approach. The third set of experiments was performed in the field.

### A. Cross-Corpus Experiments: Unseen Speakers

In a real-world setting, a network should be able to cope with unseen speech signals or speaker(s). In this set of experiments, the clean speech signals of unseen speakers, denoted by dataset 2 in Fig. 5, were considered to be unavailable to reflect a real-world setting. As shown in Fig. 5, the SSD approach could only get trained with dataset 1 for which clean speech signals were available and was then tested on dataset 2. Unlike the SSD approach, the HSD approach is capable of coping with unseen speech signals or speaker(s) by keeping the training switch *on* for unseen speaker(s) since it does not need clean speech signals for its hybrid training.

In this set of experiments, the SSD training was done for 2340 utterances of dataset 1 and the HSD training was done for 1440 utterances of dataset 1 and 900 utterances of dataset 2. More specifically, the length of the training data for both SSD and HSD remained the same in order to have a fair comparison of the two approaches. For training, the noise level was varied to generate three SNR levels of -5 dB, 0 dB, and 5 dB. Then, 360 utterances with SNR of 0 dB were used for testing.

The following experiments were conducted: (i) the network was trained with the IEEE corpus and tested with either the TIMIT or VCTK corpus; (ii) the network was trained with the TIMIT corpus and tested with the IEEE or VCTK corpus; and (iii) the network was tested with the IEEE or TIMIT corpus when the VCTK corpus was used for training. The results of these cross-corpus experiments were averaged and are reported in Table 1 as well as in Figs. 6 and 7, where $x+n$ denotes unprocessed noisy speech signals.

In these experiments, HSD had the ability to turn its training switch to *on* when untrained speech signals were encountered,

see Fig. 5. As mentioned earlier, HSD uses supervised training first to initialize the network based on public domain clean speech signals, and then uses the self-supervision approach based on noisy speech signals in the field. By turning *on* the training switch, the network associated with a noise type could continue to be trained for unseen speakers. In other words, aforementioned 1440 utterances from dataset 1 were considered for the supervised part, and 900 utterances from dataset 2 were considered for the self-supervised part. In addition to SSD and HSD, the Wiener filtering outcome is also reported here. As shown in Table 1, the performance metrics of HSD were found substantially better than those of SSD and Wiener filtering across different noise types.

The cross-corpus results when the other two datasets were used for training are provided in the form of bar charts in Figs. 6 and 7. Fig. 6 shows the results when the networks were trained on the VCTK corpus and were tested on the TIMIT or IEEE corpus. Fig. 7 reveals the results when the networks were trained on the TIMIT corpus and were tested on the IEEE or VCTK corpus. From these figures, it can be seen that the HSD approach provided superior performance over the SSD approach. For example, as shown in Fig. 6, when the network was trained on the VCTK corpus, for all the noise types, the HSD achieved better speech quality compared to the SSD approach.

Since many results are reported in this section, it helps to summarize the key findings below:

- Table 1 provides the outcome when dataset 1 in Fig. 5 was the IEEE corpus and dataset 2 was the TIMIT/VCTK corpus. The PESQ and STOI metrics for all the noise types were improved using the HSD approach compared to the SSD approach.
- Fig. 6 depicts the outcome of the developed deep speech denoising when dataset 1 was the VCTK corpus and dataset 2 was the IEEE/TIMIT corpus. Across all the noise types, HSD outperformed SSD, except for the LSD metric in the presence of wind and driving noises.
- When the TIMIT corpus was selected as dataset 1 (Fig. 7), HSD outperformed SSD across all the noise types.

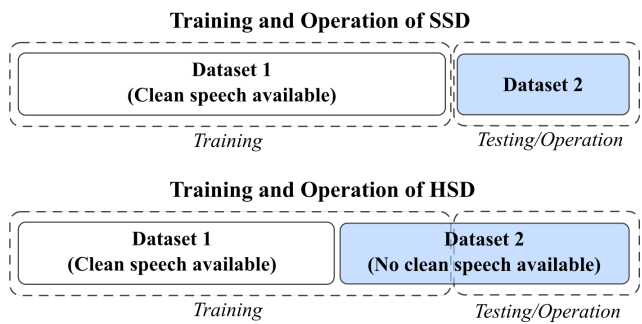

Fig. 5. Training and testing of cross-corpus experimentations.

TABLE 1. PERFORMANCE METRICS (FRAME AVERAGED ± STANDARD DEVIATION) FOR DIFFERENT NOISE TYPES WHEN THE TRAINING SET IS FROM THE IEEE CORPUS AND THE TESTING SET IS FROM THE VCTK OR TIMIT CORPUS HAVING DIFFERENT CLEAN SPEECH SIGNALS THAN THE IEEE CORPUS, THE HIGHEST VALUES ARE BOLDED.

| Noise Type | | Training from IEEE corpus | | | |
|---|---|---|---|---|---|
| | | PESQ | STOI | LSD | SSNR |
| Babble | $x + n$ | 1.37 ± 0.10 | 0.61 ± 0.04 | 1.95 ± 0.11 | -3.51 ± 1.05 |
| | Wiener | 1.39 ± 0.12 | 0.50 ± 0.05 | 1.67 ± 0.13 | -2.88 ± 0.85 |
| | SSD | 1.43 ± 0.10 | 0.62 ± 0.04 | 1.46 ± 0.12 | -0.96 ± 0.69 |
| | HSD | **1.48** ± 0.12 | **0.67** ± 0.05 | **1.29** ± 0.09 | **0.18** ± 0.61 |
| Wind | $x + n$ | 2.04 ± 0.20 | 0.85 ± 0.03 | 1.13 ± 0.08 | -2.78 ± 0.76 |
| | Wiener | 2.26 ± 0.28 | 0.71 ± 0.05 | 1.13 ± 0.06 | -2.99 ± 0.51 |
| | SSD | 2.42 ± 0.24 | 0.86 ± 0.03 | 1.13 ± 0.12 | 2.56 ± 0.31 |
| | HSD | **2.43** ± 0.24 | **0.88** ± 0.03 | **1.06** ± 0.07 | **3.19** ± 0.55 |
| Engine | $x + n$ | 1.91 ± 0.46 | 0.79 ± 0.12 | 1.37 ± 0.45 | -0.38 ± 2.71 |
| | Wiener | 1.94 ± 0.46 | 0.64 ± 0.07 | 1.61 ± 0.45 | -3.03 ± 1.68 |
| | SSD | 1.92 ± 0.43 | 0.75 ± 0.11 | 1.23 ± 0.28 | 0.74 ± 1.66 |
| | HSD | **2.11** ± 0.45 | **0.82** ± 0.11 | **1.05** ± 0.26 | **2.62** ± 2.00 |
| Driving | $x + n$ | 1.77 ± 0.21 | 0.79 ± 0.04 | 1.39 ± 0.16 | -4.02 ± 1.70 |
| | Wiener | 2.07 ± 0.21 | 0.66 ± 0.03 | 1.32 ± 0.12 | -2.78 ± 0.76 |
| | SSD | 2.03 ± 0.24 | 0.81 ± 0.04 | 1.46 ± 0.20 | 1.30 ± 0.69 |
| | HSD | **2.10** ± 0.26 | **0.84** ± 0.04 | **1.24** ± 0.14 | **2.04** ± 0.86 |

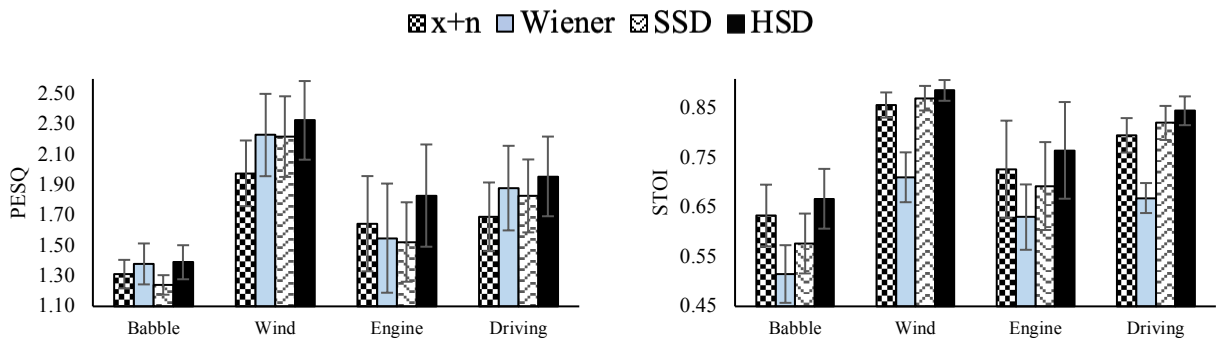

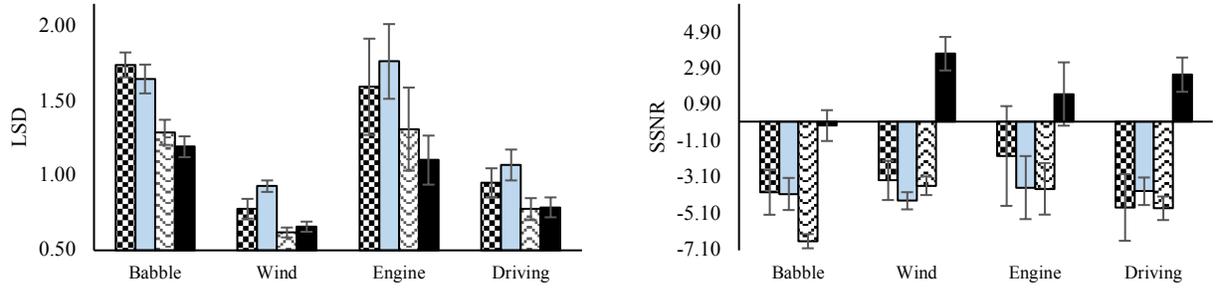

Fig. 6. Performance metrics for different noise types when the training set is from the VCTK corpus and the testing set is from the IEEE or TIMIT corpus.

In general, the results shown in the figures and tables in this section reveal that SSD failed to improve speech intelligibility compared to unprocessed noisy speech signals $x+n$ in many cases. For instance, in Fig. 6, SSD did not improve the PESQ and STOI metrics in babble and engine noises. Although the results of Wiener filtering outperformed the SSD results with respect to speech quality, Wiener filtering failed to improve speech intelligibility across the noise types compared to the raw noisy speech signals. That is why in the next set of experimentations, only the deep neural network solutions were considered for further analysis. Basically, the results reported in this section indicate that HSD is capable of performing effective speech denoising for unseen speech signals or speaker(s) while SSD or Wiener filtering are not.

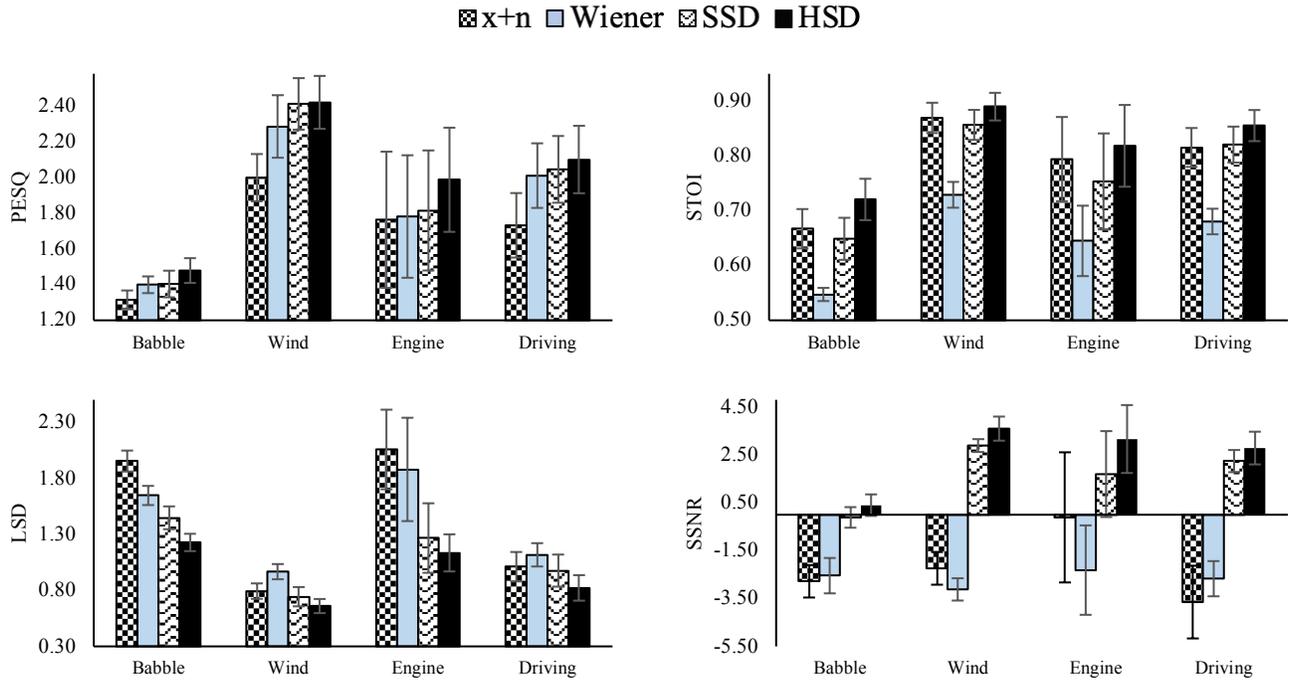

Fig. 7. Performance metrics for different noise types when the training set is from the TIMIT corpus and the testing set is from the VCTK or IEEE corpus.

### B. Cross-Corpus Experiments: Unseen SNRs

The effect of SNR variation is examined in this section. In these experiments, the trained networks were tested with unseen SNRs (3 and -3 dB) to mimic a real-world setting. Unseen SNRs refer to the mismatch between the SNRs used for training and testing. Here, the IEEE corpus was used for training and the VCTK corpus was considered to act as unseen speech signals or speakers. Similar to the previous experiments, SNRs of -5, 0, and 5 dB were used for training. The average results of these experiments across all the noise types are shown in Fig. 8 for the PESQ and STOI metrics. Other SNRs exhibited a similar behavior. As can be seen from this figure, substantial improvements in quality and intelligibility were achieved by the hybrid speech denoising.

To visually see the effectiveness of HSD, an utterance of a clean speech signal as well as its noisy version, its SSD denoised version, and its HSD denoised version are exhibited in Fig. 9. A low-frequency noise corrupted the clean speech signal in which the transient from vowel to another vowel cannot be followed visually, see Fig. 9(b). Although both SSD and HSD could reduce the noise, SSD also removed some structure of the clean speech and thus introduced speech distortion. In this figure, the two white arrows point to two regions of the spectrogram where the difference in the speech denoising is visually noticeable.

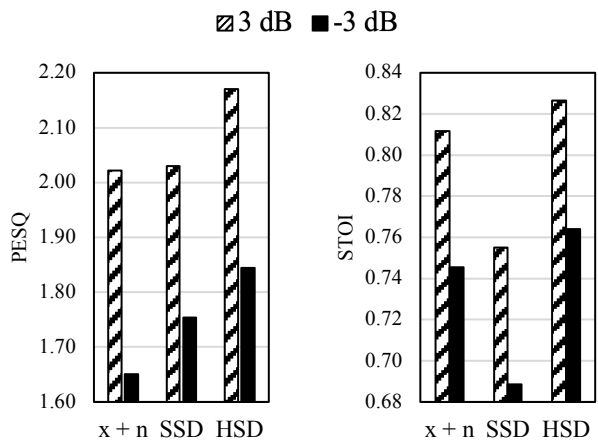

Fig. 8. Average PESQ and STOI over different noise types for unprocessed noisy speech (x+n) and denoised speech signals by SSD and HSD approaches.

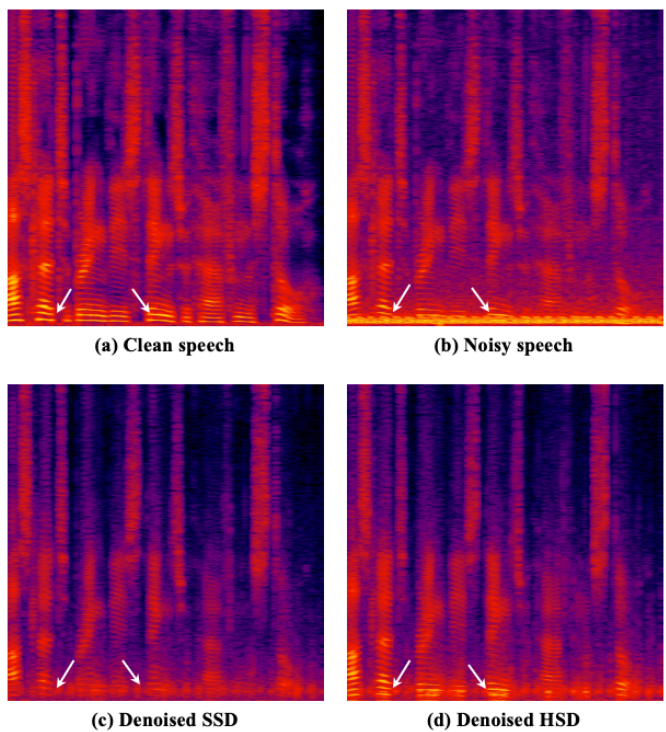

Fig. 9. Spectrograms of a sample speech signal corrupted by engine noise (a) clean speech, (b) noisy speech, (c) denoised speech by SSD, and (d) denoised speech by HSD.

### C. Field Testing

In addition to the above experiments, a field testing was conducted. This was done by using the mid-side stereo microphone Zoom iQ7 [38] connected to an iPhone allowing to capture two decorrelated signals in two channels at a sampling frequency of 48 kHz. Noisy speech signals were captured in actual audio environments or in the field for comparing the SSD and HSD approaches. The experiments performed, involved a subject reading sentences in an actual cafe environment. The recording was done for one hour.

In this set of experiments, for SSD, the network was trained by combining all the three datasets of IEEE, VCTK, and TIMIT, see Fig. 10. For HSD, 15 minutes of the two-channel field data additionally were used for training as the input and the target of the network. The spectrograms of a sample speech signal from the field and the corresponding denoised signals using SSD and HSD are shown in Fig 11. From this figure, it can be seen that under realistic conditions, SSD was not able to achieve effective denoising due to its lack of generalization capability. The last spectrogram in Fig. 11(d) shows the spectrogram of the denoised speech by HSD which appears cleaner. Considering that no clean speech signals were available when operating in the field, the objective performance metrics could not be computed and instead a portion of a sample input noisy speech signal and its corresponding SSD denoised signal and HSD denoised signal are posted at this link www.utdallas.edu/~kehtar/FieldTestingResults.html for readers to hear the superiority of the HSD approach over the SSD approach when operating in the field.

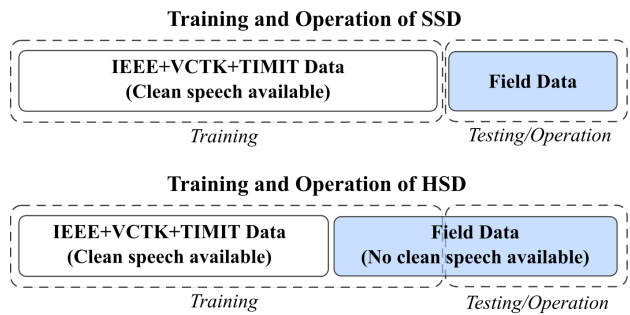

Fig. 10. Training and testing in the field.

### V. CONCLUSION

A novel hybrid deep learning-based solution has been developed in this paper for the purpose of denoising noisy speech signals in real-world audio environments by not requiring to have clean speech signals. This solution involves the use of two noisy realizations of the same speech signal as the input and the target of a fully convolutional neural network. Extensive experimentations have been conducted to compare the developed hybrid approach with the commonly used supervised approach in which clean speech signals are used as the target. The effectiveness of this approach has been established by showing that it generates improved outcomes in terms of four commonly used objective performance metrics. The developed deep speech denoising method is capable of adapting to unseen speaker(s) without the need to have clean speech signals. In summary, the developed hybrid deep speech denoising approach allows speech denoising to be carried out in the field as it does not rely on the availability of clean speech signals.

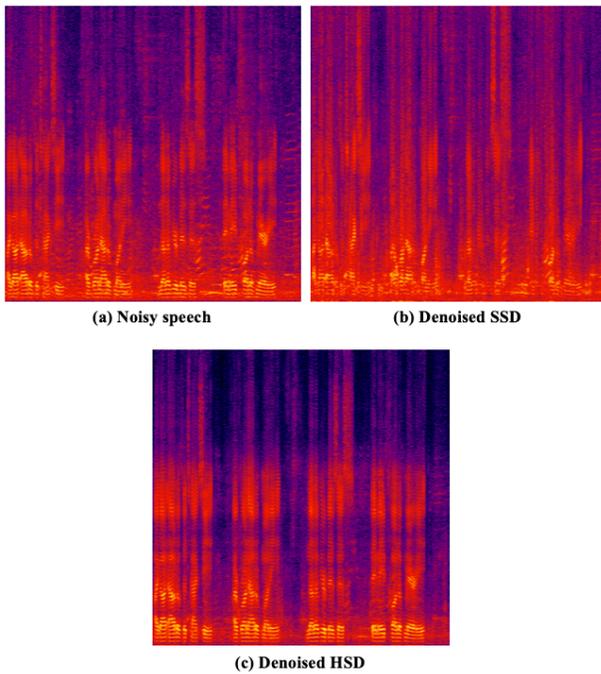

Fig. 11. Spectrograms of a sample speech signal in the field corrupted by cafeteria babble noise (a) noisy speech, (b) denoised speech by SSD, and (c) denoised speech by HSD.

## REFERENCES

[1] Berouti M, Schwartz R, Makhoul J, Enhancement of speech corrupted by acoustic noise. In: Proceedings of IEEE International Conference on Acoustics, Speech and Signal Processing, 1979;4:208-11.
[2] Sreenivas T, Kirnapure P, Codebook constrained Wiener filtering for speech enhancement. IEEE Trans. Speech Audio Process. 1996;4(5), 383–89.
[3] Alamdari, N, Yaraganalu S, and Kehtarnavaz N, A real-time personalized noise reduction smartphone app for hearing enhancement, In: Proceedings of IEEE Conference on Engineering in Medicine and Biology Society, 2018: 1-5.
[4] Xu Y, Du J, Dai L, Lee C, An experimental study on speech enhancement based on deep neural networks, IEEE Signal Process. lett., 2014;21(1):65-8.
[5] Xu Y, Du J, Dai L, Lee C, A regression approach to speech enhancement based on deep neural networks, IEEE/ACM Trans. Audio, Speech, Lang. Process., 2015;23(1):7-19.
[6] Lu X, Tsao Y, Matsuda S, Hori C, Speech enhancement based on deep denoising autoencoder, In: Proceedings of INTERSPEECH, 2013:436-40.
[7] Pascual S, Bonafonte A, Serra J, SEGAN: speech enhancement generative adversarial network, arXiv preprint arXiv:1703.09452 (2017), doi:10.21437/Interspeech.2017-1428.
[8] Kumar A, Florencio D, Speech enhancement in multiple-noise conditions using deep neural networks, arXiv preprint: 1605.02427 (2017) Available: https://arxiv.org/abs/1605.02427.
[9] Erdogan H, Hershey J, Watanabe S, Le Roux J, Deep recurrent networks for separation and recognition of single-channel speech in nonstationary background audio, in New Era for Robust Speech Recognition, 2017:165-186, Springer, Cham.
[10] Han K, He Y, Bagchi D, Fosler-Lussier E, Wang D, Deep neural network based spectral feature mapping for robust speech recognition In: Proceedings of INTERSPEECH, 2015:2484-88.
[11] Tan K, Wang D, A convolutional recurrent neural network for real-time speech enhancement, In: Proceedings of INTERSPEECH, 2018:3229-33.
[12] Martín-Doñas J, Gomez A, Gonzalez J, Peinado A, A deep learning loss function based on the perceptual evaluation of the speech quality, IEEE Signal Process. Lett., 2018;25(11):1680-84.
[13] Liu Y, Zhang H, Zhang X, Using shifted real spectrum mask as training target for supervised speech separation, In: Proceedings of INTERSPEECH, 2018, pp. 1151-1155.
[14] Zhao Y, Wang Z, Wang D, Two-stage deep learning for noisy-reverberant speech enhancement, IEEE/ACM Trans. Audio, Speech, Lang. Process., 2019; 27(1):53-62.
[15] Williamson D, Wang Y, Wang D, Complex ratio masking for joint enhancement of magnitude and phase, In: Proceedings of IEEE International Conference on Acoustics, Speech and Signal Processing, 2016:5220-24.
[16] Stowell D, Turner R, Denoising without access to clean data using a partitioned autoencoder, arXiv preprint: 1509.05982 (2015) Available: https://arxiv.org/abs/1509.05982.
[17] Furui S, Speaker-independent isolated word recognition based on emphasized spectral dynamics, In: Proceedings of IEEE International Conference on Acoustics, Speech and Signal Processing, 1986:1991-94.
[18] Purwins H, Li B, Virtanen T, Schlüter J, Chang Sh, Sainath T, Deep Learning for Audio Signal Processing, IEEE Journal of Selected Topics in Signal Processing, 2019:13(2):206-19.
[19] Rethage D, Pons J, Serra X, A Wavenet for speech denoising, In: Proceedings of IEEE International Conference on Acoustics, Speech and Signal Processing, 2018:5069-73.
[20] Germain F, Chen Q, Koltun V, Speech denoising with deep feature losses, arXiv preprint: 1806.10522v2, (2018) Available: https://arxiv.org/abs/1806.10522.
[21] Ravanelli M, Bengio Y, Speaker recognition from raw waveform with SincNet, In: Proceedings of IEEE Spoken Lang. Tech. Workshop, 2018: 1021-28.
[22] Fu S, Wang T, Tsao Y, Lu X, Kawai H, End-to-end waveform utterance enhancement for direct evaluation metrics optimization by fully convolutional neural networks, IEEE/ACM Trans. Audio, Speech, Lang. Process., 2018;26(9):1570-84.
[23] Lu Y, and Loizou P. Estimators of the magnitude-squared spectrum and methods for incorporating SNR uncertainty. IEEE Trans. Audio, Speech, and Lang. Process., 2010;19(5):1123-37.
[24] Lehtinen J, Munkberg J, Hasselgren J, Laine S, Karras T, Aittala M, Aila T, Noise2noise: Learning image restoration without clean data, arXiv preprint: 1803.04189, (2018) Available: https://arxiv.org/abs/1803.04189, 2018.
[25] Kraft S, Zölzer U, Stereo signal separation and up mixing by mid-side decomposition in the frequency-domain, In: Proceedings of 18th International Conference on Digital Audio Effects (DAFx), 2015.
[26] Kraft S, Zölzer U, Time-domain implementation of a stereo to surround sound upmix algorithm," In: Proceedings of 19th International Conference on Digital Audio Effects (DAFx), 2016.
[27] Ioffe S, Szegedy C, Batch normalization: Accelerating deep network training by reducing internal covariate shift, In: Proceedings of 32nd Int. Conf. Mach. Learn., 2015:448–56.
[28] Kingma D, Ba J, Adam: A method for stochastic optimization, arXiv preprint: 1412.6980, (2014), Available: https://arxiv.org/abs/1412.6980.
[29] Sehgal A, Kehtarnavaz N, A convolutional neural network smartphone app for real-time voice activity detection, IEEE Access, 2018;6:9017-26.
[30] Alamdari N, Kehtarnavaz N, A real-time smartphone app for unsupervised noise classification in realistic audio environments, In: Proceedings of Int. Conf. on Consumer Elec., 2019:1-5.
[31] McCloy D, Souza P, Wright R, Haywood J, Gehani N, Rudolph S, The PN/NC Corpus. Version 1.0, (2013) [Online], available: https://depts.washington.edu/phonlab/resources/pnnc/pnnc1/.
[32] DARPA-ISTO, The DARPA TIMIT Acoustic-Phonetic Continuous Speech Corpus (TIMIT), speech disc cd1- 1.1 edition, 1990.
[33] Veaux C, Yamagishi J, MacDonald K, English multi-speaker corpus for CSTR voice cloning toolkit, [Online] http://homepages.inf.ed.ac.uk/jyamagis/page3/page58/page58.html.
[34] Salamon J, Jacoby C, Bello J, A dataset and taxonomy for urban sound research, In: Proceedings of the 22nd ACM Int. Conf. on Multimedia, 2014:1041-44.
[35] Rix A, Beerends J, Hollier M, Hekstra A, Perceptual evaluation of speech quality (pesq)-a new method for speech quality assessment of telephone networks and codecs, In: Proceedings of IEEE International Conference on Acoustics, Speech and Signal Processing, 2001;2:749–52.
[36] Taal C, Hendriks R, Heusdens R, Jensen J, An algorithm for intelligibility prediction of time–frequency weighted noisy speech, IEEE/ACM Trans. Audio, Speech, Lang. Process., 2011:19(7):2125-36.
[37] Barnwell III T, Objective measures for speech quality testing J Acoust Soc Am, 1979;66(6):1658-63.
[38] Zoom iQ7, https://www.zoom-na.com/products/handy-recorder/zoom-iq7-professional-stereo-microphone-ios, 2019.